# Plasma lens for focusing attosecond pulses

E. Svirplys[1]*, H. Jones[2], G. Loisch[2], J Thomas[1], M Huck[2], O. Kornilov[1], J.M. Garland[2], J.C. Wood[2], M.J.J. Vrakking[1], J. Osterhoff[2] and B. Schütte[1]

[1] Max Born Institute for Nonlinear Optics and Short Pulse Spectroscopy; Berlin, Germany.

[2] Deutsches Elektronen-Synchrotron DESY; Hamburg, Germany

*Corresponding author. Email: svirplys@mbi-berlin.de

**Abstract:** Broadband optical pulses with attosecond to femtosecond durations provide unique opportunities for studies of time-resolved electron dynamics. However, focusing these pulses — typically ranging from the vacuum ultraviolet to the soft-X-ray region — remains challenging. Conventional refractive lenses are not suitable owing to the large dispersion and the strong absorption, while reflective optics do not suffer from these issues but have high losses. Here, we demonstrate a tunable hydrogen plasma lens to focus broadband extreme-ultraviolet attosecond pulses with energies around 20 eV and 80 eV. Simulation results suggest that stretching of the attosecond pulses is negligible, and the temporal compression is possible when atto-chirp is included. A key advantage of the plasma lens is its compatibility with nonlinear frequency conversion processes like high-harmonic generation. The different focusing properties of the fundamental and the harmonic frequencies allow for efficient separation of these components. Consequently, the transmission of high-harmonic generation beamlines can be increased to more than 80% and this approach can be suitable for applications requiring high photon flux.

**Main Text:** Refractive lenses are a straightforward method to focus light without altering its propagation direction. Their use, however, is problematic for ultrashort laser pulses with femtosecond durations. Due to dispersion, different frequency components propagate at different group velocities inside the lens, leading to pulse stretching at the focus and reducing the temporal resolution in pump-probe experiments[1]. On a microscopic level, refraction is typically governed by the interaction of light with bound electrons within the material. In response to the incident oscillating electromagnetic field, these electrons start to oscillate and emit radiation with the same frequency but with a phase delay.

When dealing with even shorter light pulses in the attosecond regime[2–6], an additional challenge for using refractive lenses arises: The spectra of these pulses typically fall within the extreme-ultraviolet (XUV) and soft-X-ray regions of the electromagnetic spectrum, where matter is highly absorbing. In the lower part of the XUV region, up to about 25 eV, this problem has been addressed through the development of gas-phase lenses[7] and metalenses[8]. Yet, like their counterparts in other spectral regions, these lenses are designed for specific wavelength ranges and are not suitable for focusing extremely short light pulses in the attosecond regime. Instead, mirrors are commonly used to focus attosecond pulses, though they come with several disadvantages: Mirrors suffer from low reflectivities, rapid degradation effects[9], and may require sophisticated alignment strategies[10,11].

## Plasma lens

Here we propose and demonstrate a refractive plasma lens for focusing broadband XUV pulses with an initial duration in the attosecond range. The refractive index $n$ originating from the interaction of an electromagnetic field with free electrons is given by[12]

$$n = \sqrt{1 - \omega_p^2/\omega^2}, \quad (1)$$
$$\omega_p^2 = n_e e^2/\varepsilon_0 m_e. \quad (2)$$

Here $\omega_p$ is the plasma frequency, and $\omega$ is the angular frequency of the external electromagnetic field. The former depends on the free-electron density $n_e$, the electron charge $e$, the permittivity in free space $\varepsilon_0$, and the electron mass $m_e$. For frequencies exceeding the plasma frequency, the refractive index has real values below unity, leading to a phase advance of the incident electromagnetic field. To exploit these properties of the propagation of light in a free electron density for focusing attosecond light pulses, a concave radial electron density profile with a minimum electron density on the optical axis is required. Optimal performance is achieved for a parabolic electron density profile[13,14]. Unlike conventional optics, the plasma lens is not affected by XUV-induced damage, as the plasma can be replenished with every laser shot. In this context, plasma lenses have also been proposed for focusing high-power laser pulses[15–17].

The plasma lens is ideally suited for focusing attosecond pulses comprising a broad spectral bandwidth. While the refractive index varies across the spectrum, pulse stretching within the plasma lens under our experimental conditions ($\omega_p$ = 0.56 PHz and photon energy of 80 eV) is small. Moreover, the negative dispersion of the plasma offers the possibility to temporally compress positively chirped attosecond pulses, as will be discussed later.

To create conditions suitable for focusing broadband XUV pulses, we generated a plasma from hydrogen molecules using a capillary discharge[18–21] (Fig. 1). Due to the low electron binding energies, hydrogen can be nearly fully ionized[13,18]. The absence of bound electrons results in an exceptionally high transmission of the XUV pulses, which is an attractive feature of plasma that sets it apart from all other forms of matter.

The capillary discharge source consists of a sapphire block that contains a 5-cm-long, 300-µm-diameter channel. Hydrogen is supplied via four inlets and ionized by applying a discharge current pulse using electrodes attached at both ends of the capillary (see Methods and SI for details). Following ignition of the plasma, its temperature distribution becomes inhomogeneous due to energy exchange with the relatively cooler capillary walls, resulting in the formation of a parabolic plasma density profile[13,22]. An attractive feature of the plasma lens is the ability to control its focal length by varying the hydrogen pressure inside the capillary.

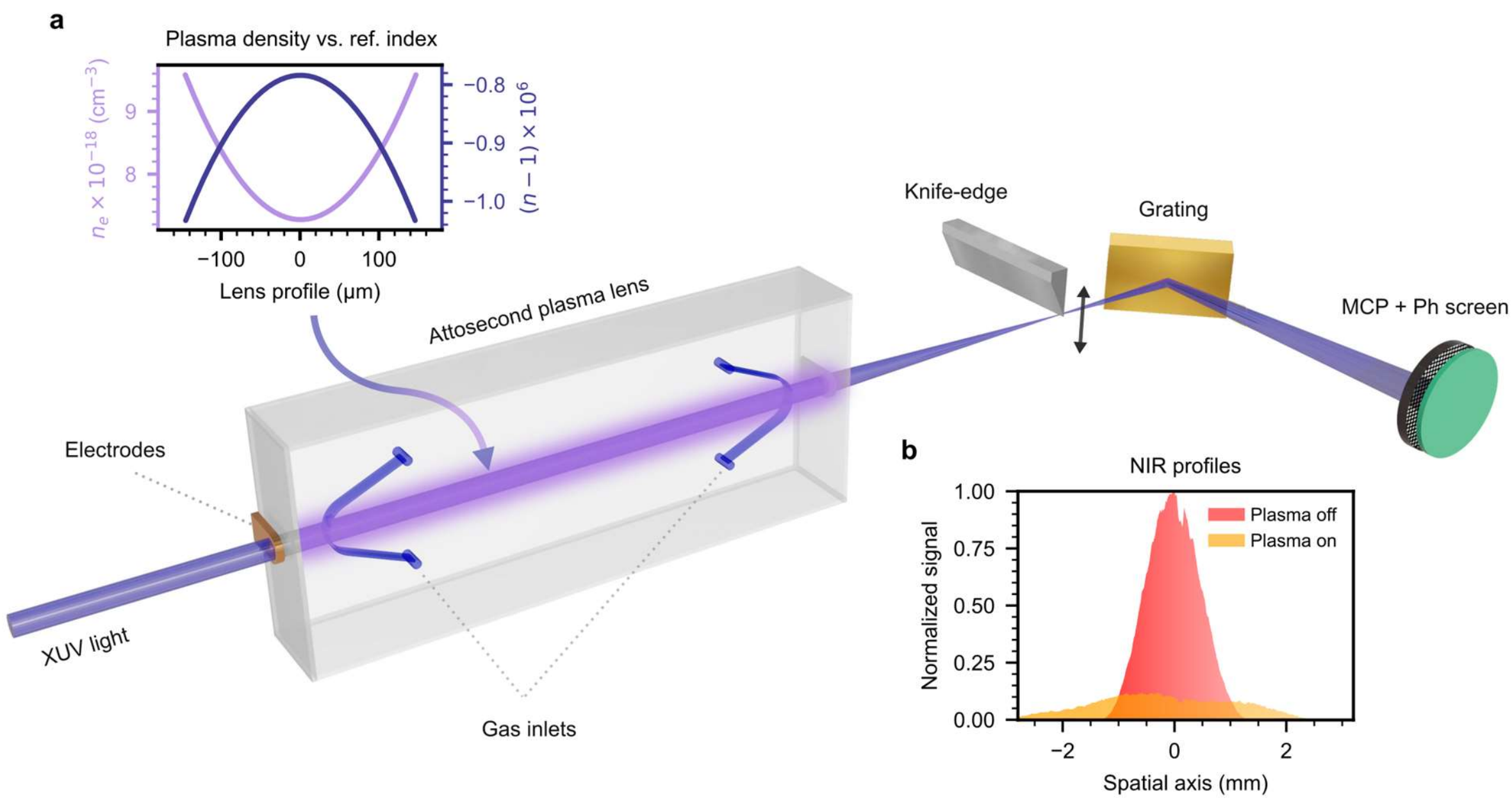

**Fig. 1 Scheme of the plasma lens experiment**. Capillary discharge source consisting of a sapphire block that has a 5-cm-long, 300-µm-diameter channel. Hydrogen is delivered via four symmetric inlets. Copper electrodes are attached on both ends, applying a current pulse to ignite a plasma. An XUV pulse produced by high-harmonic generation (HHG) is transmitted through the channel, and a knife edge is placed at the focal plane to measure the XUV focus size. Following diffraction from a grating, the XUV spectrum is recorded using a microchannel plate (MCP) / phosphor screen assembly, and the data are acquired using a CCD camera. **a,** Illustrative parabolic plasma density profile, $n_e$ (purple curve), across the capillary profile following plasma cooling at the channel walls. The corresponding refractive index, $n$, for 80 eV XUV light is shown in blue. **b**, NIR beam profiles in the XUV focal plane without (red curve) and with plasma (yellow curve), showing that the NIR beam is effectively defocused after the plasma lens.

## Experimental demonstration of focusing broadband XUV pulses

The experiments were performed utilizing an 18-m-long beamline that was optimized for the generation of intense attosecond pulses via HHG[23,24] (see Methods for details). Following spectral filtering using an ultrathin Sn foil, XUV pulses centered at 20 eV were focused by the plasma lens, which was mounted at a distance of 13 m from the HHG target. To characterize the focus, a knife edge, placed 20 cm after the lens, was scanned across the XUV beam profile. The XUV beam was diffracted from a grating onto a microchannel plate (MCP) / phosphor screen assembly (see Fig. 1). Measurements were conducted during the peak plasma transmission window, approximately 160 ns after the discharge (see SI). Fig. 2**a** shows the spatial profiles of the XUV beam for an evacuated capillary (yellow triangles) and following plasma generation using a hydrogen pressure of 63 mbar (purple circles). Focusing of the XUV beam is clearly observed, characterized by a decrease of the $1/e^2$ radius from 97±2 μm to 40±2 μm. The focused radius as a function of photon energy and hydrogen pressure is depicted in Fig. 2**b**, showing good agreement with numerical results obtained using a wave propagation method (dashed line, see SI for details).

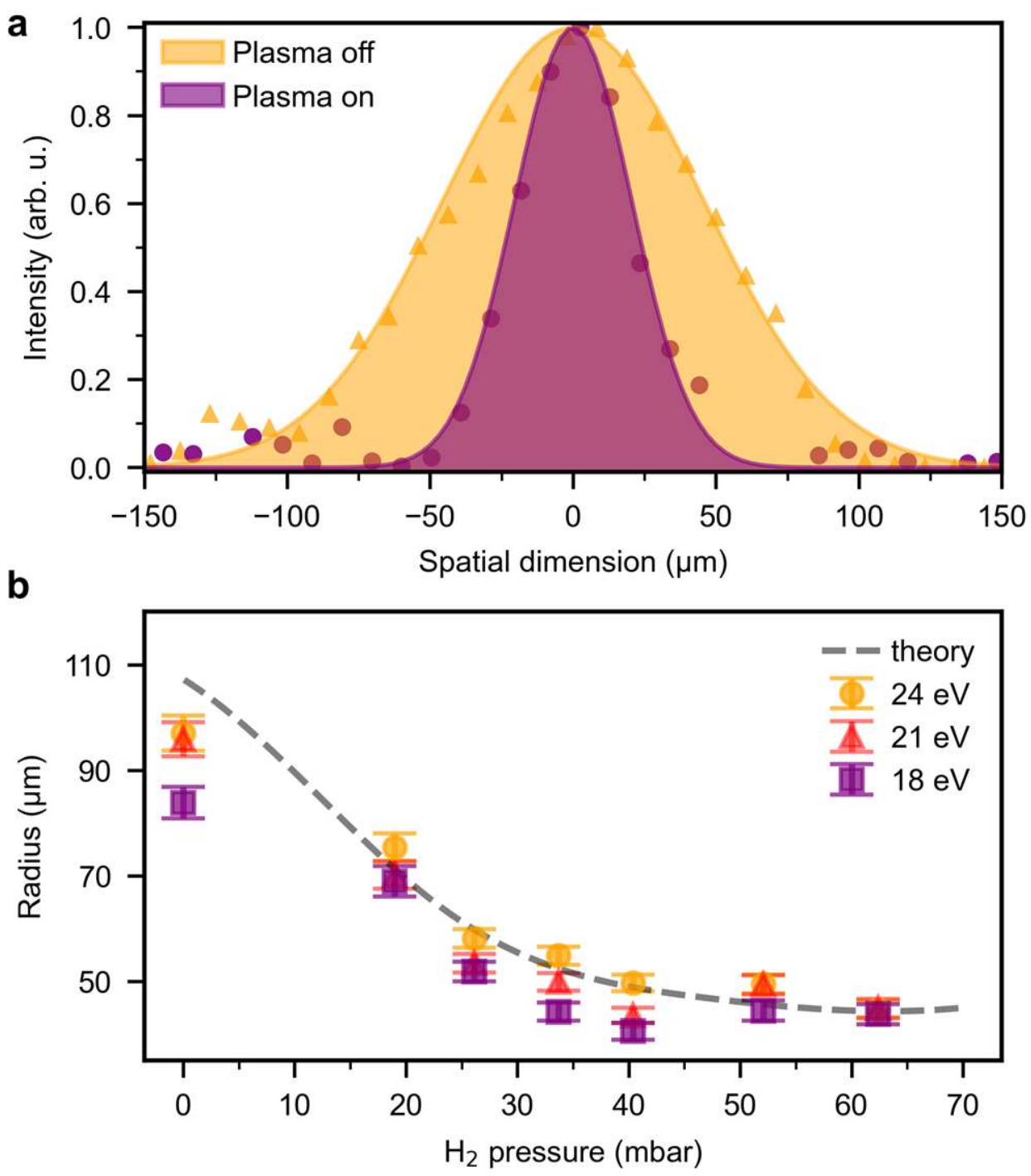

**Fig. 2. Experimental realization of the plasma lens**. **a**, XUV beam profiles measured 20 cm after the lens following ignition of a plasma using a hydrogen pressure of 63 mbar (purple circles) and for an evacuated capillary (yellow triangles). Focusing from a radius of 97±2 μm to 40±2 μm is observed. These values were obtained from fitting a Gaussian function (solid curves) to the derivative of the recorded transmission data, thereby recovering the beam's intensity profile. Observed transmission during plasma operation was approximately 62%, indicating that hydrogen was not fully ionized. **b**, Spectrally resolved fitted XUV beam radii as a function of the hydrogen pressure inside the capillary, with error bars indicating the standard deviation of the fitted values. The solid curve shows the simulated XUV beam waist, calculated for a photon energy of 21 eV (see SI).

While focusing of XUV beams in the 20-25 eV region using lenses has been demonstrated previously[7,8], no lenses have so far been developed for higher XUV photon energies. XUV light around 70-100 eV (corresponding to wavelengths of 12.4-17.7 nm) is of particular interest, as it enables the interaction with core electrons in atoms and molecules[25,26]. Furthermore, EUV lithography, a critical technology for fabricating integrated circuits with structure sizes in the few-nanometer range, relies on light at a wavelength of 13.5 nm[27,28].

We have generated broadband XUV pulses centered at 80 eV (Fig. 3**a**) using HHG in Ne and a Zr foil for spectral filtering. Since the refractivity decreases with increasing photon energies for otherwise identical conditions (Eq. 1), this configuration results in a longer XUV focal length. The XUV focus size was measured directly on an MCP / phosphor screen assembly that was placed 120 cm behind the lens, see Fig. 3**c**. The pressure-dependent XUV beam radius is presented in Fig. 3**b**, showing good agreement between the experiment and the simulation. The relatively large XUV focus waist of 80±1 µm is a result of the small numerical aperture used in this experiment. Fig. 3**d** presents a spectrally resolved simulation of the XUV beam profile under the same conditions. The beam waist radius integrated over the entire bandwidth is only slightly larger than the beam waist radius at 80 eV (80 µm compared to 79 µm), showing that detrimental effects due to chromatic aberrations are small.

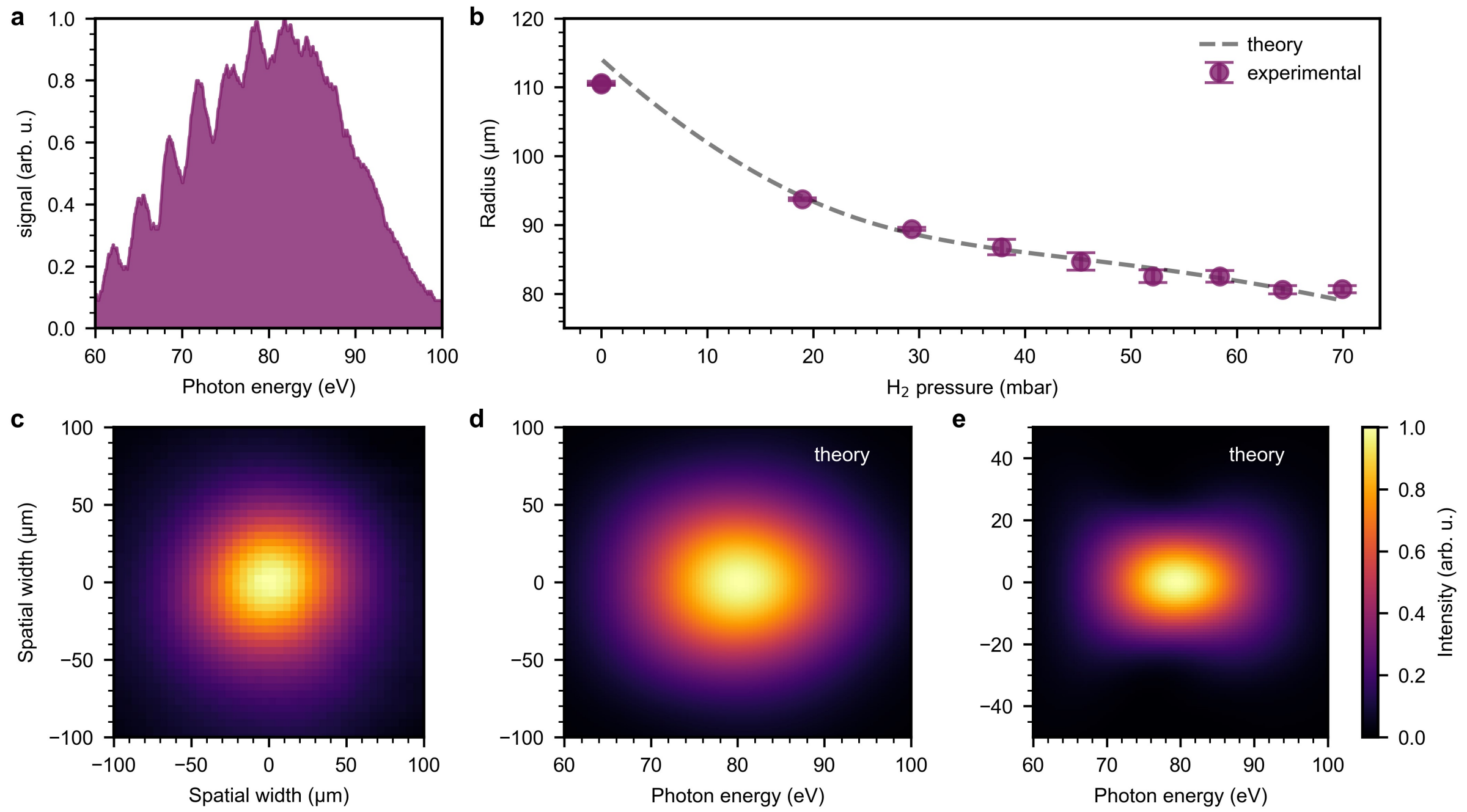

**Fig. 3 Focusing XUV pulses at 80 eV. a**, XUV spectrum obtained from HHG in Ne measured after a Zr filter. **b**, Comparison of the experimentally obtained XUV beam radii (violet circles) and the simulated XUV beam radii (dashed line) as a function of the hydrogen pressure. The error bars were calculated in the same way as in Fig. 2b. **c**, XUV beam profile measured in the presence of plasma at a hydrogen pressure of 70 mbar. **d**, Simulated spectrally resolved XUV beam profile for the same experimental conditions as in **c**, showing a waist of 80 µm. **e**, Spectrally resolved XUV beam profile, calculated for a capillary with a length of 10 cm and a pressure of 200 mbar, showing a waist of 27 µm.

Tighter focusing of the XUV beam can be achieved in the future by increasing the length of the plasma lens. Our simulations suggest that for a capillary length of 10 cm, an XUV beam waist radius of 27 µm is achievable when using a hydrogen pressure of 200 mbar (Fig. 4**e**). In this case, the focal length is reduced to 41 cm. Detrimental effects due to chromatic aberrations remain small in this case as well: the waist at 80 eV is 23 µm, compared to 27 µm when integrated over the full spectrum. Importantly, discharge capillaries initially filled with hydrogen at pressures of up to 330 mbar have already been demonstrated[29]. Additionally, even stronger focusing may be achievable in the future by employing laser-heated capillaries, which further increase the on-axis plasma temperature and steepen the refractive index gradient across the capillary profile[30–32].

**Temporal properties of attosecond pulses focused by a plasma lens**

To study the temporal properties of attosecond pulses focused by the plasma lens, we have performed pulse propagation simulations, see SI. First, we considered a transform-limited attosecond pulse with a central photon energy of 80 eV and a full width at half maximum (FWHM) duration of 90 as passing through a plasma lens filled with 70 mbar of hydrogen. As shown in Fig. 4**a**, the pulse experiences negligible stretching to 96 as in the XUV focal plane. Our analysis indicates that the group delay dispersion (GDD) is the main contribution to pulse stretching (see SI).

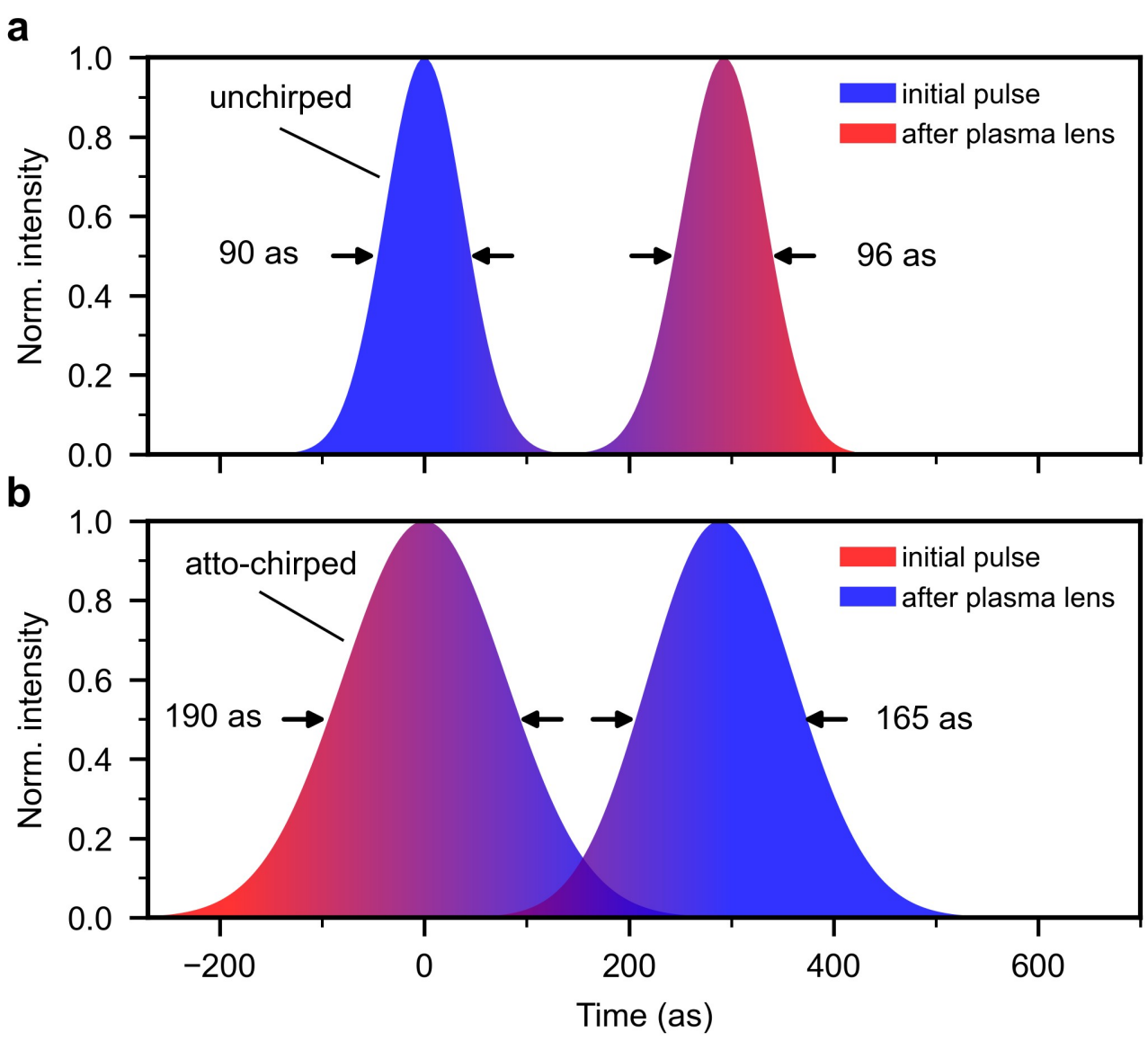

**Fig. 4. Simulated temporal profiles of attosecond pulses focused by a plasma lens. a**, Temporal profiles of a transform-limited 90 as pulse centered at 80 eV before (blue) and after (violet) passing through a plasma lens filled with 70 mbar of hydrogen, resulting in negligible stretching to 96 as. **b**, Simulation of an initially chirped 190 attosecond pulse propagating through the same plasma lens, demonstrating compression to 165 as. This suggests that a plasma lens can be used for atto-chirp compensation.

However, when accounting for the positive chirp (known as the atto-chirp)[33] acquired by attosecond pulses during high-harmonic generation (HHG), the hydrogen plasma may be used to temporally compress the pulses[34]. Under typical experimental conditions, our simulations show that pulses with an initial duration of 190 as may be compressed to 165 as (Fig. 4**b**). At a higher pressure of 200 mbar, further compression down to 127 as is anticipated (see SI), suggesting that the attosecond plasma lens could enhance the temporal resolution in pump-probe experiments.

## Discussion

The properties of the plasma lens may be exploited in the future to substantially increase the transmission of HHG beamlines. Following XUV pulse generation via HHG, the attosecond pulses co-propagate with the more powerful NIR driving pulses. To avoid strong NIR laser fields being present in the focal plane, thin metal filters are typically used to attenuate the NIR power. However, this approach comes with significant XUV transmission losses. A key advantage of the plasma lens is that the focusing properties of NIR and XUV beams are entirely different. The focal plane of the NIR beam is located inside the capillary, and the NIR beam is divergent after exiting the plasma lens, as shown in Fig. 1**b.** As a result, an estimated NIR intensity of only $10^8$ W/cm$^2$ is obtained at the XUV focal plane (see SI). This value is negligible in typical pump-probe experiments. Consequently, the use of thin metal filters may become unnecessary in the future. In addition, we benefit from the low XUV absorption within the lens. The plasma lens enables a beamline transmission exceeding 80 %, compared to typical values on the order of 10-20 %[35,36]. This provides novel opportunities for photon-demanding attosecond applications, such as applications in attochemistry using molecules that can only be supplied at low densities[37–39].

Another drawback of thin metal filters is their low damage threshold, limiting their use to sufficiently low NIR intensities. To attenuate the NIR power, additional mirrors may be employed[40], but this further reduces the XUV beam transmission. Alternatively, the filter must be positioned sufficiently far from the HHG source. By removing the need for a filter, a plasma lens can be placed much closer to the HHG source, enabling a significantly more compact setup.

For the proof-of-principle experiments described here, the capillary discharge source was operated at 20 Hz. Capillary discharge sources operating at kHz repetition rates have already been demonstrated[41], which could enhance the XUV photon flux available in experiments and may thereby enable temporal characterization of the focused XUV pulses using attosecond streaking or RABBITT[3,4,42]. Alternatively, lasers may be used for plasma generation at kHz repetition rates. Laser-produced plasmas with suitable characteristics have already been demonstrated for waveguiding applications[43–46], where axicon lenses were used to create extended plasma channels with minimal electron density along the optical axis.

The plasma lens is suitable for focusing light at other wavelengths, including few-femtosecond deep-ultraviolet (DUV) and vacuum-ultraviolet (VUV) pulses, which can be efficiently generated in gas cells[47] or through soliton dynamics in hollow-core fibers[48–52]. An important challenge is attenuation of the fundamental light, which can be addressed using mirrors at Brewster's angle[50]. However, this approach suffers from significant reflection losses and mirror degradation. These obstacles can be overcome by using a plasma lens, which promises a high transmission and effective defocusing of the driving laser. We have performed simulations confirming that the plasma lens can efficiently separate the fundamental NIR beam from the co-propagating VUV pulse, allowing for effective isolation of the VUV light (see SI for details).

In conclusion, we have proposed and demonstrated a lens that focuses broadband XUV pulses by the interaction with free electrons in a nearly fully ionized plasma. The plasma lens is characterized by high transmission, immunity to XUV laser damage, and inherent wavelength tunability.

## Methods

### *Attosecond pulse generation*

NIR driving laser pulses were obtained from a Ti:Sapphire laser system operating at 1 kHz (Spitfire, Spectra Physics). In the current experiment, 37-fs-long pulses centered at 800 nm with a pulse energy of 9 mJ were used. To enable the generation of near-isolated attosecond pulses, the NIR pulses were temporally compressed down to 3.7 fs using a three-stage post-compression scheme, resulting in a pulse energy up to 4.9 mJ[23]. Attosecond pulses were generated using an 18-m-long high-harmonic generation (HHG) beamline[23,24]. To this end, the NIR pulses were focused using a telescope consisting of a convex ($f$ = -1 m) and a concave ($f$ = 0.75 m) spherical mirror. HHG was performed in a 4-cm-long cell that was statically filled with gas. Argon was used for the generation of harmonics around 20 eV and 60 eV, while neon was used for the generation of harmonics around 80 eV. The experimental vacuum chamber, in which the attosecond plasma lens was mounted, was placed approximately 13 m downstream from the HHG cell. The experimental setup is presented in Fig. S1. To select specific spectral regions of the XUV pulses, either a 100 nm Sn or a 150 nm Zr foil was inserted. The measured XUV spectra are shown in Fig. S2. For the knife-edge measurements conducted 20 cm after the capillary (Fig. 2), an attosecond pulse filtered by a Sn filter was used (Fig. S2**a**), resulting in a full-width-half-maximum (FWHM) of 6.8 eV. The XUV spectrum centered at 80 eV, used in Fig. 3, is shown in Fig. S2**b**. Additionally, the dashed line in Fig. S2**b** represents an XUV spectrum with a FWHM of 22 eV, which was used for the calculations shown in Fig. 3 and Fig. 4. Transmission data for both filters were obtained from Henke et al[53].

### *Plasma lens pressure calibration*

A constant hydrogen gas flow into the capillary was achieved using a thermal mass flow controller (Bronkhorst). A series of XUV light transmission measurements were performed to accurately determine the relationship between the hydrogen mass flow and the resulting gas density inside the capillary. To this end, the spectrally resolved transmission of the XUV light through the capillary was measured as a function of the hydrogen flow (see Fig. S3**a**). By applying Beer-Lambert's law and using known molecular hydrogen cross-sections ($\sigma$)[54], the neutral atomic hydrogen density $n_a$ inside the capillary was determined according to

$$n_a = -\frac{2\ln(T)}{\sigma L},$$

where $T$ is the measured transmission of a specific harmonic of XUV light, and $L$ is the length of the capillary. To calculate the pressure from the gas density, the ideal gas law was used, assuming a temperature of 295 K. Calculated hydrogen pressures are presented in Fig. S3**b**. The measured data points were fitted with a power function of the form $ax^b$, where $a$ and $b$ are the fitting coefficients and $x$ is the gas flow. The resulting calibration fit is displayed in Fig. S3**b** as a red curve.

### *Hydrogen ignition*

The hydrogen gas inside the capillary was ignited using a high-voltage current pulser, which generates sub-µs-long electron pulses with a 500 A peak current at a repetition rate of up to 1 kHz. The current pulse was delivered to the lens through copper electrodes attached to the ends of the capillary, and the discharge current was monitored using an oscilloscope (see shaded area in Fig. S4). For the experiments presented in the main text, the pulser repetition rate was set to 20 Hz.

**Acknowledgments:** We thank Simon Hooker for fruitful discussions during the initial project conceptualization, Roman Peslin for his technical assistance, Christoph Reiter for providing laser support, and Melanie Krause for implementing safety protocols in the laboratory. ES would like to thank Augustė Mažeikaitė for help with three-dimensional graphical visualization.

**Funding:**

The project was funded by the Deutsche Forschungsgemeinschaft (DFG, German Research Foundation) – project number 456137830.

**Author contributions:**

Conceptualization: BS, OK, MJJV

Resources: GL, MJG, JCW, JO, OK

Investigation: ES, HJ, JMG, JT, BS

Formal analysis: ES, HJ, MH

Data curation: ES,HJ

Visualization: ES

Funding acquisition: BS, OK, JO

Project administration: BS, JO, MJJV

Writing – original draft: ES, BS, MJJV

Writing – review & editing: ES, HJ, JT, GL, MH, JCW, OK, MJJV, JO, BS

**Competing interests:** Authors declare that they have no competing interests.

**Data and materials availability:** All data used in the main text or the supplementary information are accessible on Zenodo[55].

## S1. Comparison between a toroidal mirror and a plasma lens

It is useful to briefly discuss the most standard scheme currently used to focus attosecond pulses, which utilizes a toroidal mirror in a 2f-2f geometry. In this case, the XUV source is imaged to the experimental region without any (de-)magnification. This step is necessary to attenuate the NIR power, which is orders of magnitude higher than the XUV power. To achieve this, thin metal filters are typically used (sometimes in combination with mirrors at Brewster's angle or with an antireflective coating for the IR pulses). These optics must be placed sufficiently far from the HHG source to prevent damage. Since the XUV focus size depends on the IR focus size used for HHG, it typically falls within the range of several tens of micrometers. This value is sufficient for most attosecond experiments that utilize an XUV-NIR pump-probe scheme.

When compared to a typical setup making use of a toroidal mirror, we have achieved a similar focus size using a plasma lens, but we benefit from the following advantages: (i) Since no metal filter is needed, a much higher transmission can be obtained. (ii) Alignment is straightforward. (iii) The plasma lens can be placed closer to the HHG source, making experimental setups more compact. (iv) The propagation direction of the XUV pulses is maintained, leading to more flexible experimental setups.

## S2. Experimental setup for attosecond pulse generation

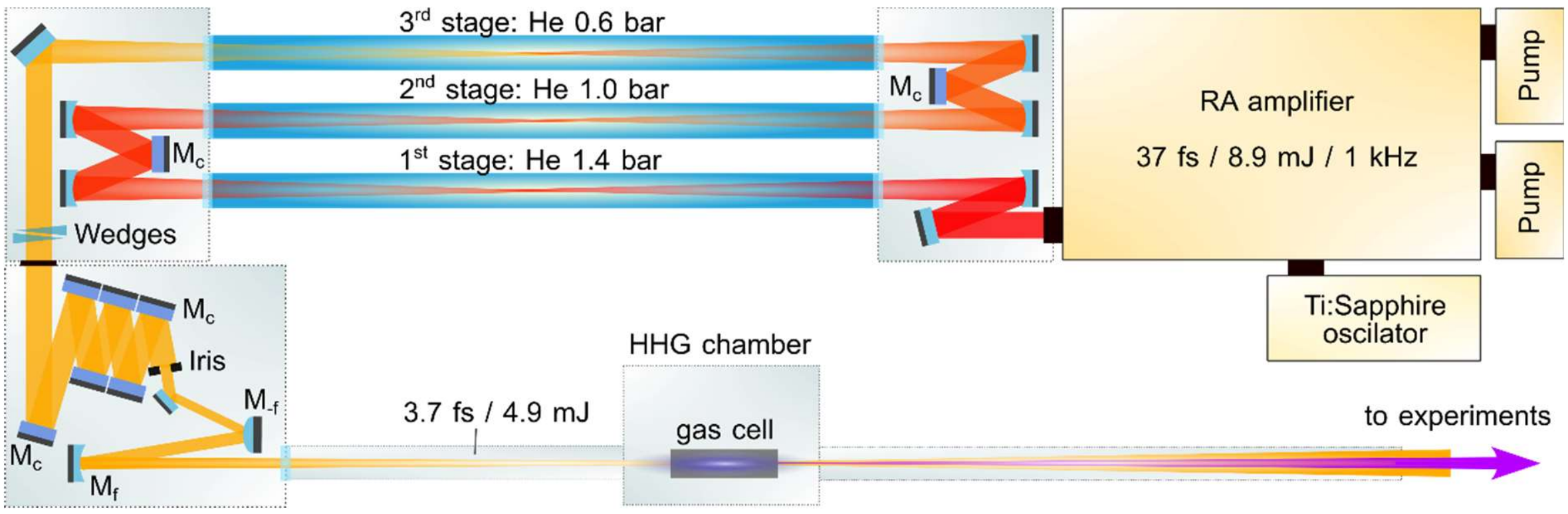

**Fig. S1. Laser setup for attosecond pulse generation.** The setup consists of a Ti:sapphire laser system, which generates 37-fs-long NIR pulses. These pulses were temporally compressed using a cascaded post-compression technique in three helium-filled tubes at varying pressures. Chirped mirrors ($M_c$) were used to temporally compress the pulses after each stage. The compressed NIR pulses were focused by a telescope consisting of convex ($M_{-f}$) and concave ($M_f$) mirrors. Attosecond pulses were generated in a gas cell. The experimental chamber was placed 13 m downstream from the HHG cell.

## S3. Spectra of attosecond pulses

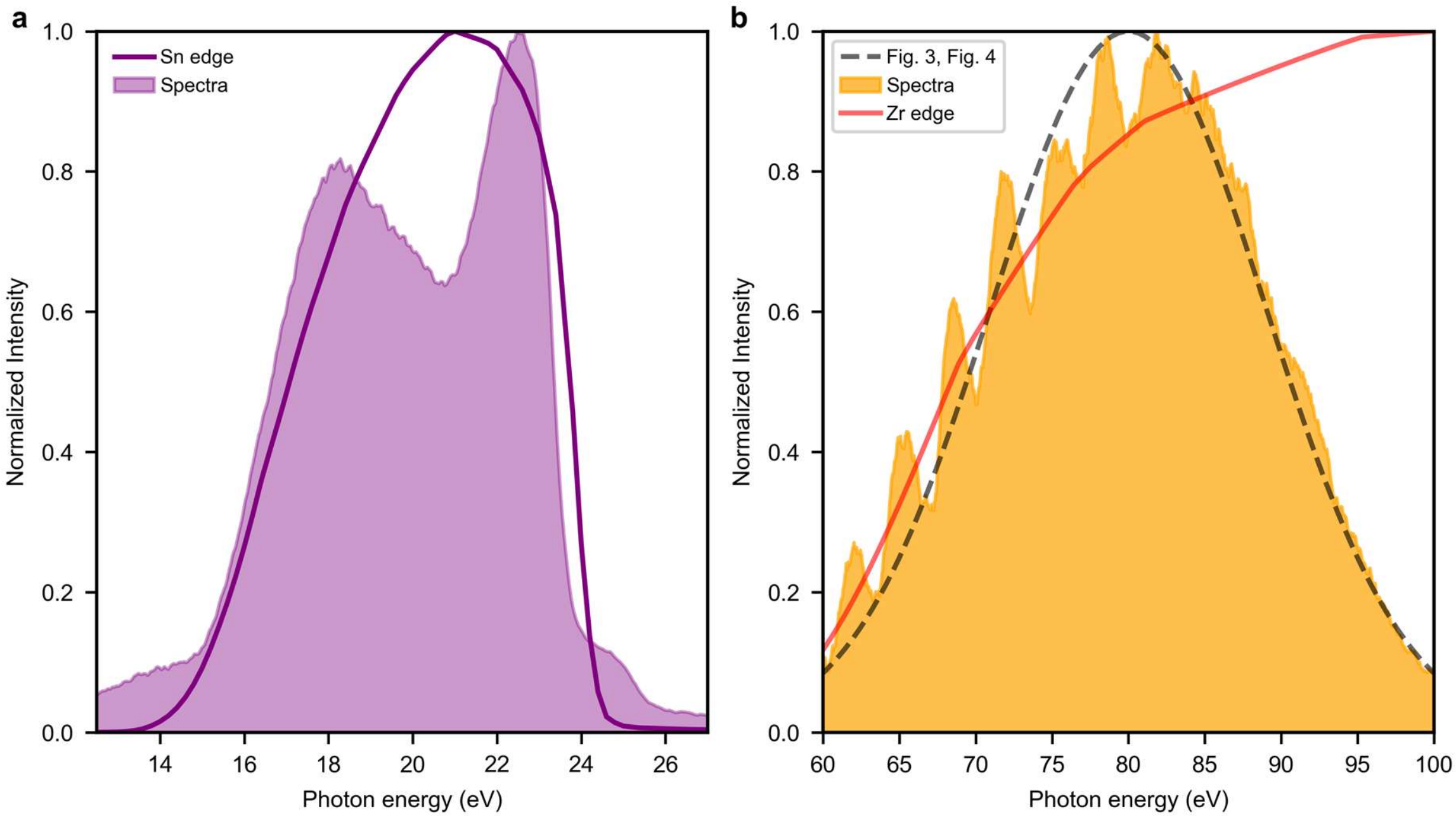

**Fig. S2. Measured XUV spectra. a**, XUV spectrum used for the experiment shown in Fig. 2 of the main manuscript (shaded area). The solid line shows the normalized transmission window of the Sn foil used in the experiment. **b**, XUV spectrum used for the experiment shown in Fig. 3 of the main manuscript (shaded area). The solid line shows the transmission curve of the Zr filter. The dashed line shows the Gaussian-shaped XUV spectrum that was used for the calculations in Fig. 3 and Fig. 4.

## S4. Plasma lens pressure calibration

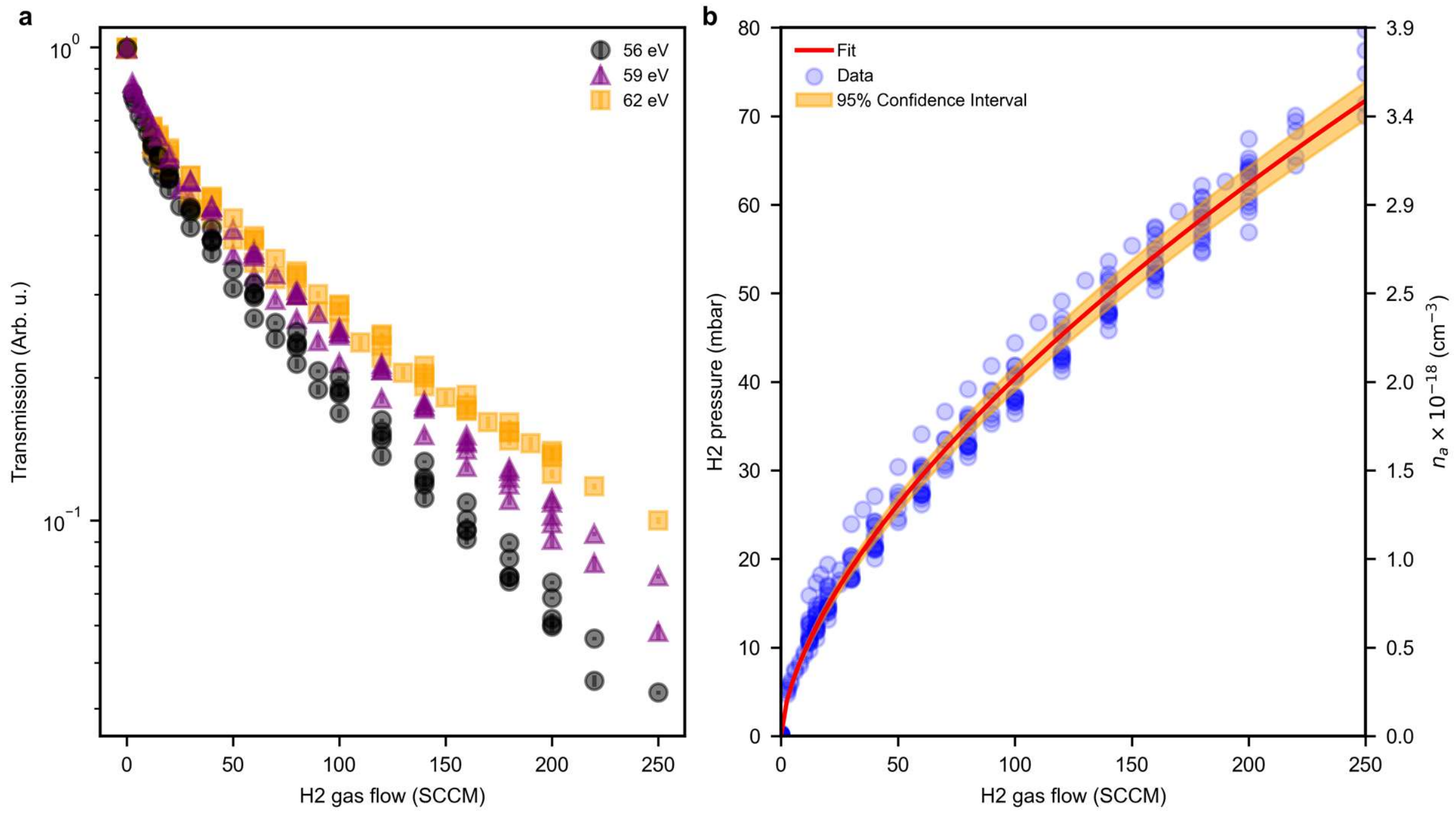

**Fig. S3. Calibration of the hydrogen pressure. a**, XUV light transmission through the capillary as a function of photon energy and hydrogen mass flow into the capillary, expressed in standard cubic centimeters per minute (SCCM). **b**, Calculated hydrogen pressure as a function of the mass flow (circles). The solid line represents sum-of-squares fit applied to the measured data (sample size of 263), and the shaded area indicates the 95% confidence interval of the fit. The right axis shows the calculated initial atomic hydrogen pressure, $n_a$.

## S5. Plasma lens transmission

The transmission of XUV light through the plasma lens depends on the ionization fraction of hydrogen. Capillary discharge sources typically reach peak ionization within the first few hundred nanoseconds after the discharge and subsequently recombine to a neutral state over many microseconds due to energy loss through heat exchange with the capillary walls[1]. Therefore, the precise timing of the discharge current relative to the XUV pulse is important to achieve maximum transmission. Fig. S4 shows the measured XUV transmission through the capillary as a function of the delay between the initiation of the discharge current and the arrival time of the XUV pulse at the capillary. For an XUV pulse centered at 80 eV, the capillary was filled with 63 mbar of hydrogen, and the transmission was calculated relative to the XUV flux measured after the capillary when it was evacuated. A maximum transmission of 84% was observed approximately 160 ns after initiating the discharge. This was also the timing used for focusing experiments presented in the main text. The shaded area represents the normalized measured discharge current trace.

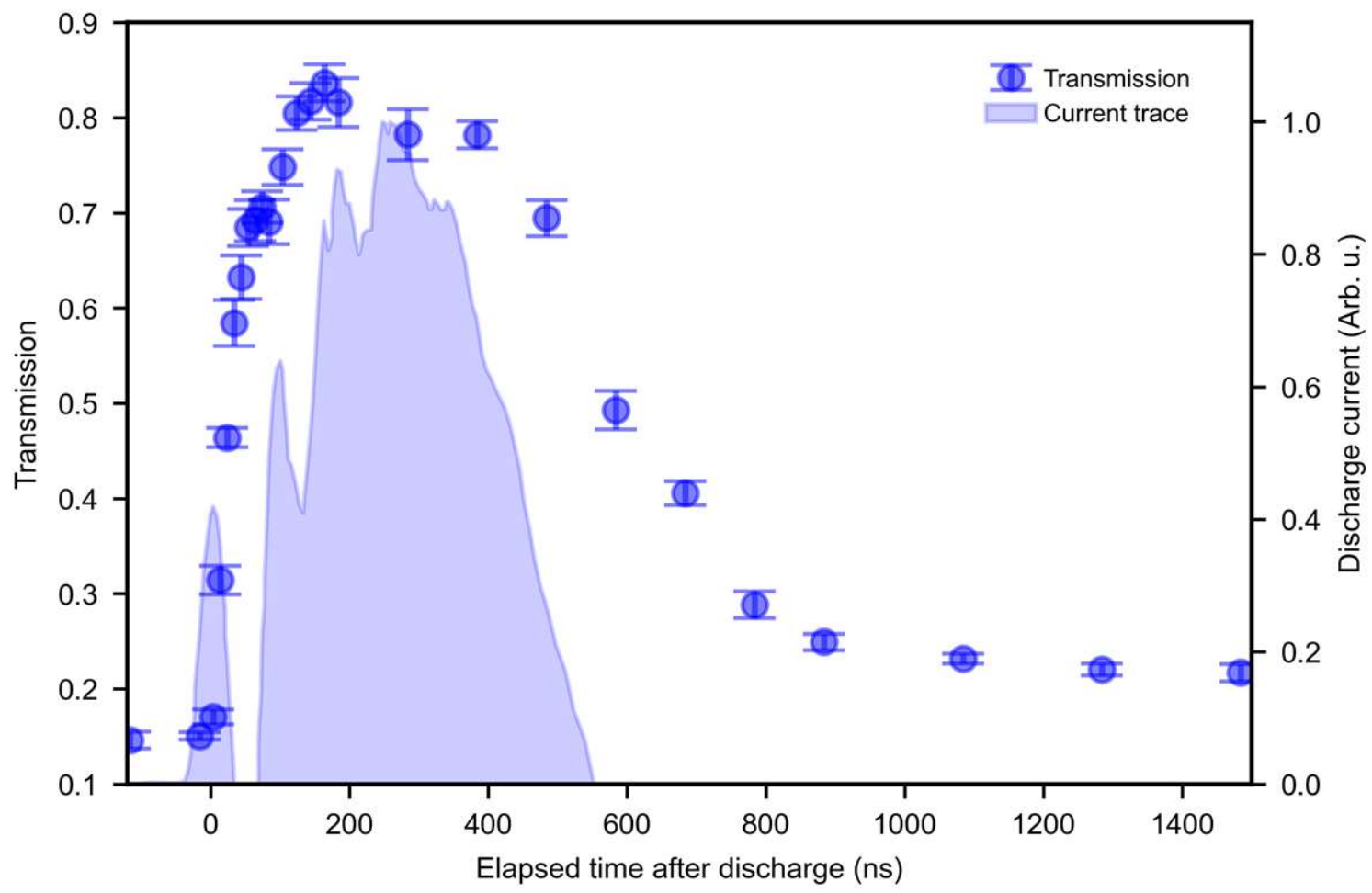

**Fig. S4. Plasma lens transmission.** XUV light transmission through the capillary filled with 63 mbar of hydrogen as a function of time after the discharge (circles). Each point represents the mean value of 10 measurements of the transmitted XUV flux, with error bars indicating the standard deviation. The shaded area shows the recorded discharge current trace.

## S6 Longitudinally resolved plasma density measurement

For optimal focusing operation of the plasma lens, the plasma profile should be parabolic in the transverse direction and homogeneous in the longitudinal direction. In this context, the plasma density was measured along the capillary length using plasma emission spectroscopy[2].

A plasma emission spectrometer consisting of an optical imaging system and a grating was used, enabling imaging of a spectrally resolved side-view of the capillary (see Fig. S5**b**) using a temporally gated CCD camera. This setup recorded the spatially resolved plasma density evolution with a temporal resolution down to 20 ns. The first spectral line in the Balmer series, Hα (656 nm), in hydrogen has a well-known spectral line broadening temperature and density dependence, which allowed for direct extraction of the plasma temperature and density from the measured broadening of this spectral line[2].

In Fig. S5**a**, the evolution of the spatially resolved plasma density is shown. The plasma density was measured for ten spatial regions along the capillary length. Using an initial pressure estimated as 60 mbar, an upper on-average plasma density up to 2 x $10^{18}$ $cm^{-3}$ was measured in the time interval of 0.1–0.15 µs after the discharge. Furthermore, the plasma density remained fairly uniform along the longitudinal capillary axis during this time interval, with a calculated mean relative error of 10% across all spatial regions. A significant deviation from this uniformity was observed for times exceeding one microsecond as the gas is expelled from the capillary through the ends.

a

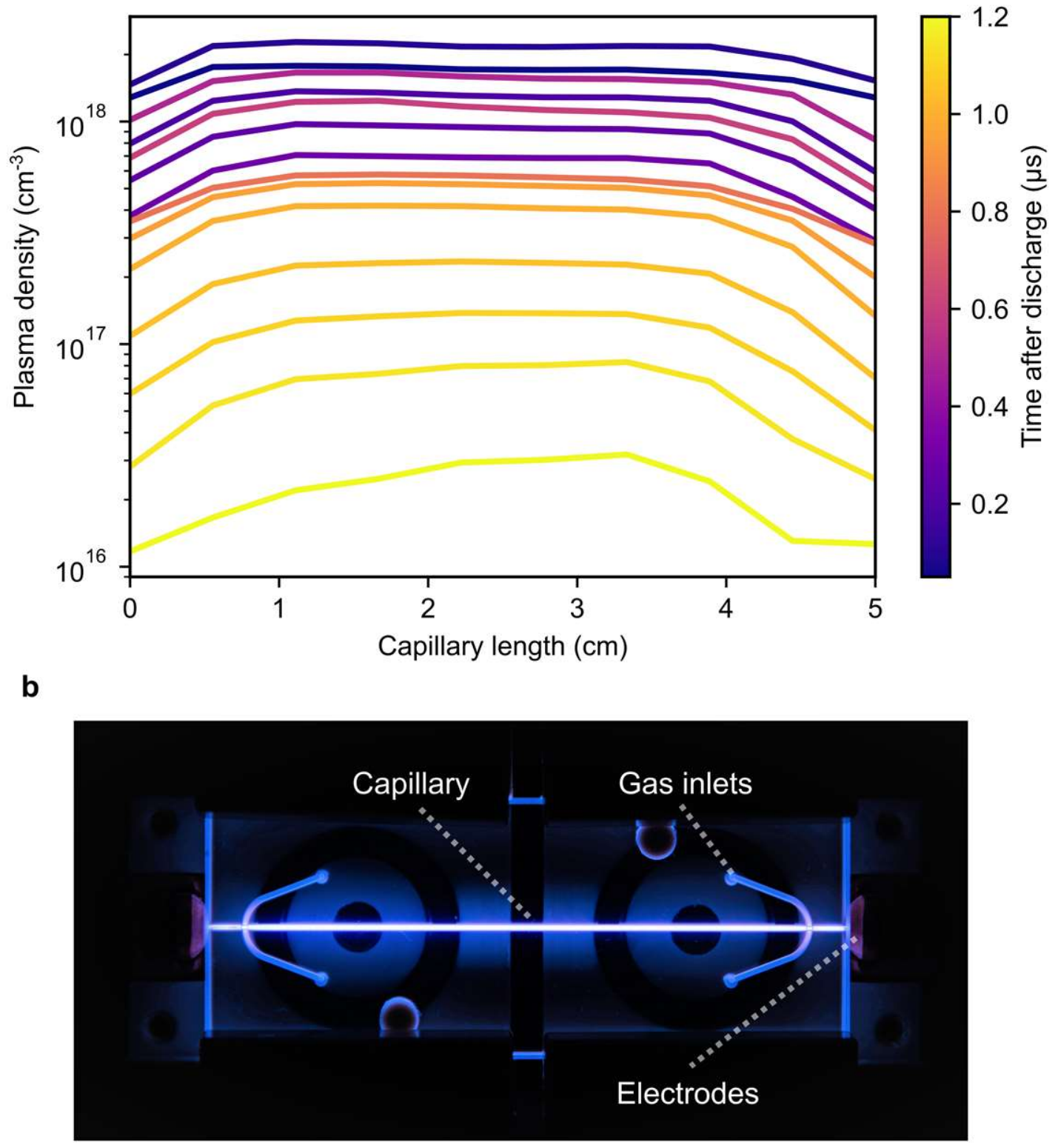

**Fig. S5. Longitudinally resolved plasma density measurement. a**, Evolution of the longitudinally resolved plasma density inside the capillary discharge source following ignition. The initial gas pressure inside the capillary was set to 60 mbar. **b**, Side-view of the plasma lens during plasma emission.

## S7 Plasma density profile model

To calculate the XUV beam profile at the focus of the plasma lens, we have assumed a parabolic plasma density profile and a Gaussian XUV beam profile at the entrance of the capillary. To calculate the curvature and magnitude of the parabolic plasma density profile, an analytical formula retrieved from magneto-hydrodynamics calculations was used, as derived for a hydrogen-filled capillary discharge source similar to ours[1]. This formula can be used to calculate the plasma density inside the capillary for times exceeding 100 ns following the discharge, when the plasma distribution is predominantly governed by electron conduction. The only variables needed for calculating the plasma density distribution $n_e(r)$ inside our plasma lens using this formula are the capillary radius $r_{ch}$ and the initial atomic hydrogen density $n_a$:

$$n_e\,(r) = n_{eo}\left(1 + 0.33\frac{r^2}{{r_{ch}}^2}\right), \tag{S1}$$

$$n_{e0} = 0.7364 \cdot n_a. \tag{S2}$$

Here, $r$ is the radial distance from the capillary center and $n_{eo}$ is the on-axis electron density. The refractive index profile can then be calculated using Eqs. (1) and (2). Examples of the calculated plasma densities and the refractive index along the center of the capillary are shown in Fig. S6.

It is worth noting that plasma densities inside a capillary discharge source are generally not perfectly parabolic across the entire capillary profile, as density gradients are typically steeper near the capillary walls[3]. However, the theoretical framework presented below, based on a parabolic density approximation, yields predictions that closely match the measured focus (see Figs. 2**b** and 4), supporting the validity of this model.

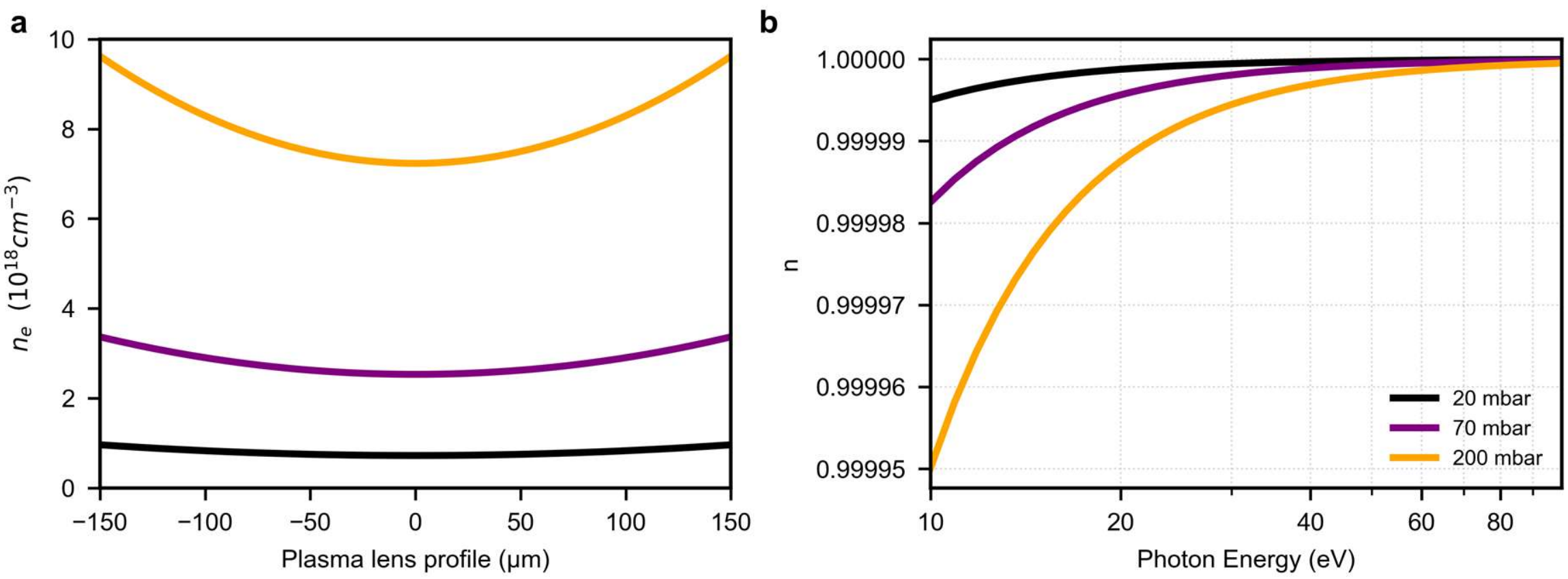

**Fig. S6. Calculated plasma densities and refractive index. a**, Electron density as a function of the transverse spatial coordinate for three different values of the initial hydrogen pressure, which was calculated using Eq. S2. **b**, Refractive index as a function of the photon energy for hydrogen pressures of 20 mbar, 80 mbar, and 200 mbar. These calculations were performed using Eqs. (1) and (2).

## S8 Numerical wave propagation simulation

We have used a numerical wave propagation method (WPM)[4,5] to simulate the propagation of an XUV beam through our plasma lens. In this two-dimensional approach, the *r* direction represents the transverse coordinate, while the $z$ direction corresponds to the propagation axis. The incident Gaussian field is defined as:

$$u(r,z) = \left(\frac{\omega_0}{\omega_z}\right) e^{-\left(\frac{r}{\omega_z}\right)^2} e^{ik_0\frac{r^2}{2R}}, \tag{S3}$$

where $\omega_0$ and $\omega_z$ denote the beam radius at the waist and position $z$, respectively, $k_0$ is the vacuum wave number, and $R$ represents the beam curvature. The WPM numerically solves the Helmholtz equation using the split-step Fourier method, modeling the field evolution $u(r,z)$ at small increments $dz$ in the propagation direction:

$$u(r, z + \mathrm{dz}) = \mathcal{F}^{-1}\left[e^{ik_z dz}\mathcal{F}[u(r,z)]\right], \tag{S4}$$

where $\mathcal{F}$ and $\mathcal{F}^{-1}$ denote the Fourier and inverse Fourier transforms, and

$$k_z = \sqrt{nk_0 - k_r}. \tag{S5}$$

Here, $n$ is the refractive index that can be calculated from the model presented above, and $k_r$ is the transverse wave vector component corresponding to the spatial frequency along the *r*-axis.

An example of this numerical wave propagation through the plasma lens is shown in Fig. S7. The complex refractive index distribution used in the WPM is shown in Fig. S7**a**, where the capillary walls were modeled using a complex refractive index $\tilde{n}$ with a real part of $n = 0.966$ and an imaginary part $\kappa = 0.0413$ as an optical extinction coefficient[6]. The intersection region between the capillary walls and the plasma was modeled using an apodizing function, which smooths the transition and reduces artifacts in the calculation. The effect of the plasma lens on a Gaussian beam at 80 eV is demonstrated by comparing the propagation of a beam through an evacuated capillary (Fig. S7**b**) and a capillary filled with 200 mbar of hydrogen (Fig. S7**c**). The refraction induced by the plasma modifies the beam trajectory, confirming the focusing effect.

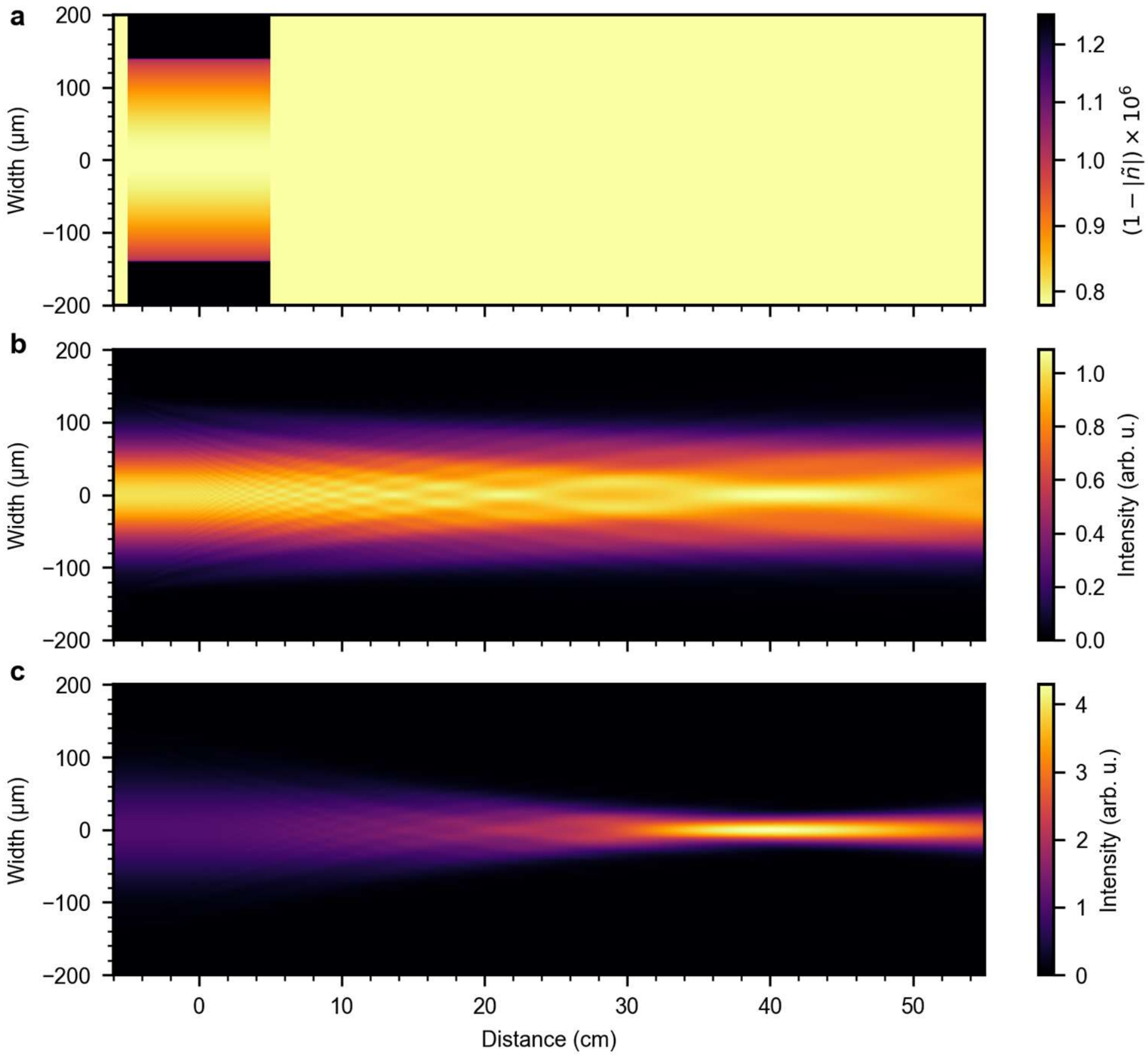

**Fig. S7. Numerical wave propagation of an 80 eV Gaussian beam inside and outside a 10 cm-long plasma lens. a**, Real contribution to the refractive index as a function of the transverse and longitudinal directions, as calculated using the model provided above and Eqs. (1)-(2). The black-shaded area represents the capillary walls. **b**, Gaussian beam propagation inside and after an evacuated capillary. **c**, Propagation of the same beam when assuming 200 mbar of discharge ionized hydrogen gas inside the capillary.

## S9 NIR pulse divergence after the plasma lens

As discussed in the main text, the plasma lens not only focuses the XUV beam, but also substantially reduces the fundamental NIR field strength in an experiment, due to the significant divergence of the NIR beam.

To investigate this effect, we used the same NIR pulse employed for attosecond pulse generation and measured its divergence after passing through the plasma lens. A CCD camera was positioned approximately 40 cm after the capillary to capture the NIR beam profile at various hydrogen pressures inside the capillary. The measured profiles were spatially integrated, and the radius at the $1/e^2$ intensity level was calculated. The resulting NIR beam radii (red circles) as a function of hydrogen pressure are presented in Fig. S8**a**, illustrating a large divergence.

Theoretically calculated radii (blue curve) are also shown in Fig. S8**a**, showing good qualitative agreement with the experimental data. Due to the broad spectrum of the NIR pulse, numerical wave propagation through the plasma lens was performed for each wavelength within the NIR spectrum. The data were then combined, taking into account the quantum efficiency of the CCD camera.

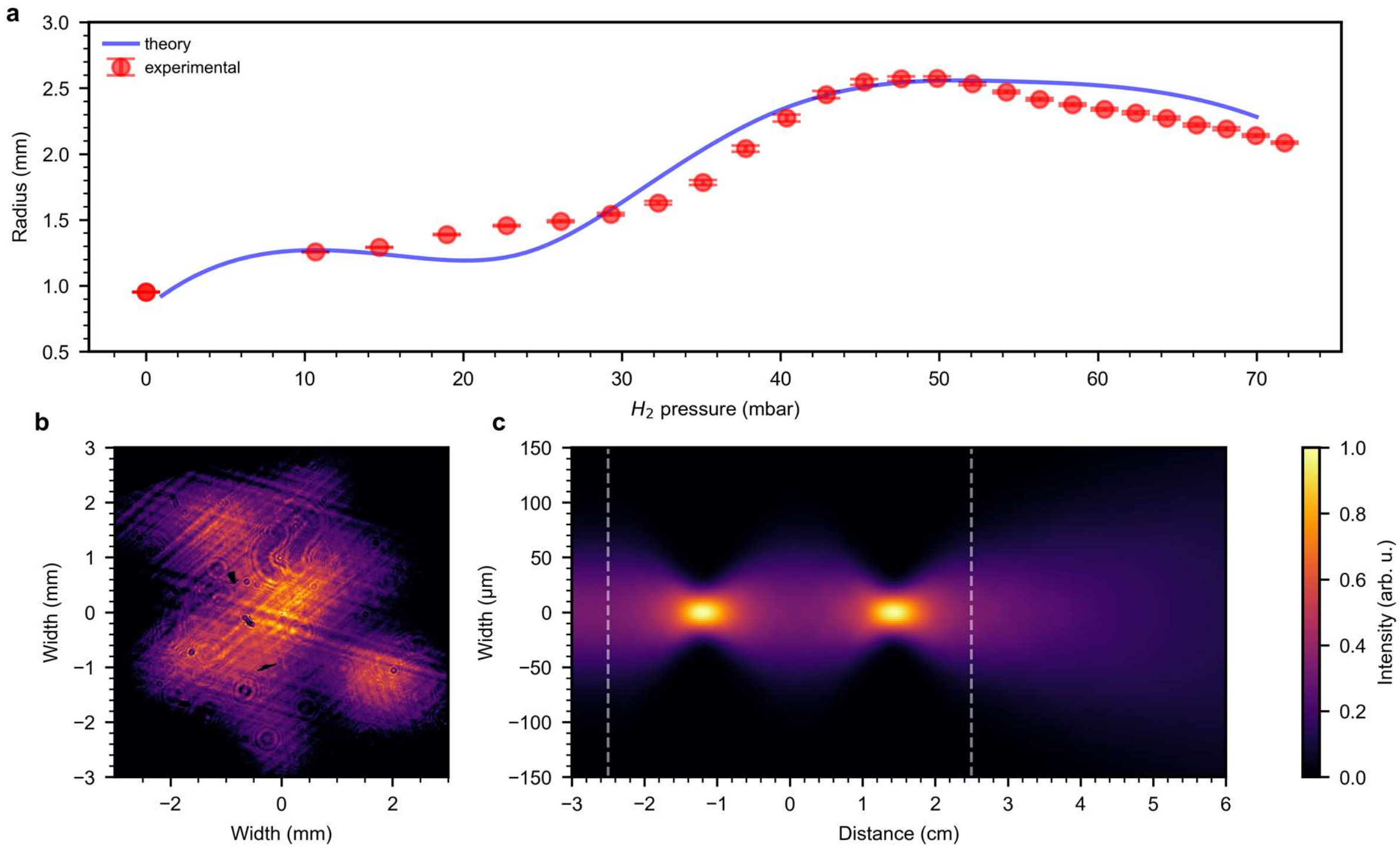

**Fig. S8. NIR pulse divergence after the plasma lens. a**, Comparison of the measured NIR beam radius (circles) at a distance of 40 cm after the lens and the computed NIR beam radius (solid curve) as a function of the hydrogen pressure. Each experimental point is given by the mean of Gaussian-fitted radii from 4 measured NIR beam profiles, with error bars representing the standard deviation of the fitted values. **b**, NIR beam profile measured after the plasma lens at a hydrogen pressure of 50 mbar. **c**, Simulated NIR beam propagation through a 5 cm-long capillary filled with 50 mbar of hydrogen, showing that the NIR beam is focused inside the capillary. The dashed vertical lines indicate boundaries of the capillary.

The NIR beam radius at a distance of 40 cm from the lens is 2.5 mm when applying a hydrogen pressure of 50 mbar (Fig. S8**b**). At the focal plane of the XUV beam, which is at a distance of 1.2 m from the capillary (see Fig. 3), the NIR beam radius is expected to be 7.5 mm, which is about 90 times larger than the measured XUV waist. We further benefit from the fact that the divergence of the NIR driving laser pulses is about one order of magnitude larger than the divergence of the generated harmonics. As a result, an NIR pulse energy of less than 10 μJ was transmitted by the capillary, i.e. 1% of the energy. Taking into account that the initial 3.7-fs-long NIR pulse, is stretched to about 50 fs due to plasma dispersion, we estimate the NIR pulse peak intensity at the XUV focal plane to be on the order of $10^8$ W/cm$^2$. This value can be neglected in most attosecond experiments.

Fig. S8**a** shows that the NIR beam waist decreases slightly from its maximum value at pressures exceeding 50 mbar. This behavior can be understood by examining the numerical wave calculations for 800 nm light propagating through a plasma capillary filled with 50 mbar of hydrogen, as shown in Fig. S8**c**. The simulations indicate that the NIR beam is focused twice within the capillary. Consequently, the beam divergence outside the capillary is influenced by these internal focal points, leading to a reduced divergence at higher than 50 mbar hydrogen pressures. However, if the pressure would be even further increased above 70 mbar, the divergence would begin to increase again.

This oscillatory behavior is strongly influenced by the initial beam size and is associated with a matched spot-size parameter, which is commonly used to characterize NIR waveguiding through a capillary discharge source[1,3,7]. When the initial beam size is equal to the matched spot size, the beam radius within the capillary remains nearly constant along the length of the capillary.

## S10 VUV propagation through the plasma lens

As discussed in the main text, we suggest that the plasma lens may be used for vacuum-ultraviolet (VUV) pulses generated in gas cells or hollow-core fibers. To demonstrate this capability, we performed numerical simulations using the wave propagation method for a 2-cm-long, 600-µm-wide capillary filled with 60 mbar of hydrogen.

Fig. S9 illustrates the distinct effects of the plasma lens on NIR (800 nm) and VUV (165 nm) light. The NIR beam is sharply focused near the exit of the capillary, leading to significant divergence beyond this point (Fig. S9**a**). In contrast, the VUV beam is focused at a distance of 20 cm after the lens (Fig. S9**b**). Figure S9**c** compares the beam waist evolution of both wavelengths, showing that at a distance of 20 cm after the lens, the NIR beam has expanded to a radius of 2.5 mm, while the VUV beam remains tightly confined with a radius of just 64 µm, which is 39 times smaller, allowing for effective separation between the two wavelengths.

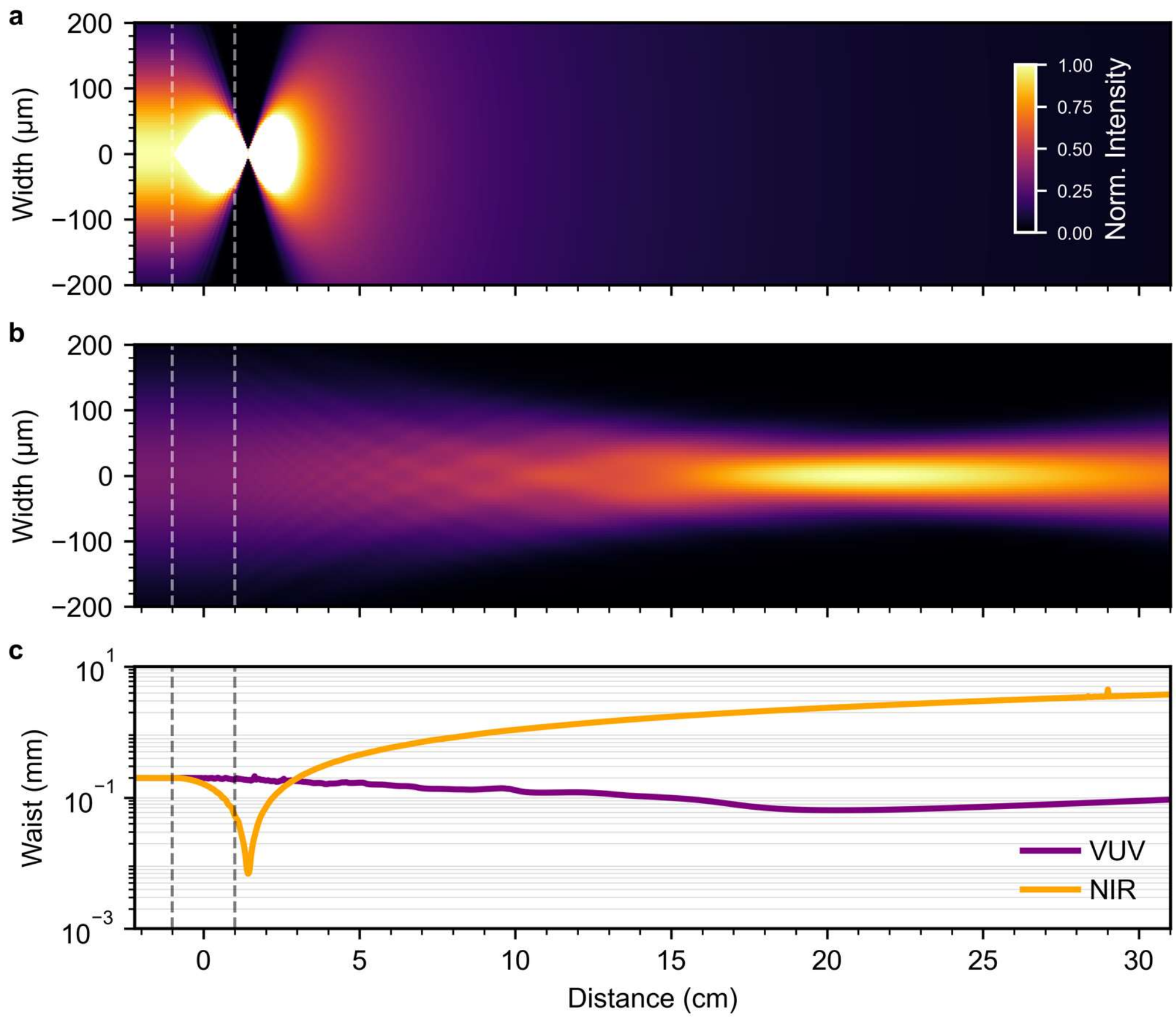

**Fig. S9. Comparison of the NIR and VUV beam divergence after the plasma lens. a**, Sharp focusing of the NIR beam near the lens exit, leading to a substantial divergence afterwards. **b**, Focusing of the 7.5 eV VUV beam at approximately 20 cm after the lens. **c**, Comparison of the NIR and VUV beam waists at various positions before and after the lens, showing that the NIR beam size is 39 times larger than the VUV beam at 20 cm after the lens. The dashed vertical lines indicate plasma lens capillary boundaries. All calculations were performed using the WPM described above, with a capillary length of 2 cm, a diameter of 600 µm, and a pressure of 60 mbar.

## S11 Simulations of the temporal properties of an attosecond pulse focused by the plasma lens

The simulation of the temporal properties follows the formalism provided in Ref.[8] and has been adapted for this study. Similar to the calculations described above for the WPM, we used a two-dimensional (*x,z*) grid with a refractive index $n(x)$ given by Eqs. (1) and (S1). The initial electric field distribution was described by a Gaussian pulse envelope, $u_0$, as defined in Eq. (S3). We employed the eikonal approximation to determine the angular frequency $\omega$ dependent phase and the amplitude of the pulse after propagation through the lens:

$$u(x, z = 0) = u_0(x)e^{i\omega L[n(x)-1]/c}, \tag{S6}$$

where $L$ is the length of the lens capillary and $c$ is the speed of light. After passing through the lens, the pulse propagation in free space was modeled using the small-angle approximation. The spatial distribution of the attosecond pulse frequency components at the focal position $f$ is then given by:

$$\tilde{u}(x, z = f) = \int u(x)e^{i\frac{\omega x^2}{2fc}}dx. \tag{S7}$$

Then, the spatio-temporal intensity distribution of the attosecond pulse was obtained by taking the square of the inverse Fourier transform of Eq. (S7):

$$U(x, t) = |\mathcal{F}^{-1}[\tilde{u}(x, w)]|^2. \tag{S8}$$

Eq. (S8) can then be used to calculate the pulse envelope at the focal plane by integrating over the spatial axis.

To account for the positive chirp in the attosecond pulses originating from HHG, a spectral phase term $\Phi(\omega)$ was added to the initial electric field distribution expressed as:

$$\Phi(\omega) = \frac{1}{2}C(\omega - \omega_0)^2. \tag{S9}$$

where $C$ is the chirp parameter with units in as$^2$, and $\omega_0$ is the central frequency of the envelope.

Simulation results for a plasma lens filled with 200 mbar are shown in Fig. S10. Figs. S10**a-c** present the spatiotemporal properties of a transform-limited (unchirped) 90-attosecond-long pulse centered at 80 eV. Fig. S10**a** shows the spatiotemporal profile of the pulse before entering the plasma lens. After propagation through the lens (Fig. S10**b**), the pulse is focused and experiences small temporal broadening to 117 as. This is more clearly seen in Fig. S10**c**, where spatially integrated temporal envelopes of the initial and final pulse are shown.

Figures S10**d**–**f** shows the same calculations but for the case of an initially positively chirped (atto-chirped) 189 as pulse. The input spatiotemporal structure (Fig. S10**d**) evolves significantly as it passes through the plasma lens (Fig. S10**e**), resulting in temporal compression to 127 as. The integrated temporal envelopes for the chirped case are shown in Fig. S10**f**, illustrating the effectiveness of the plasma lens in compensating atto-chirp.

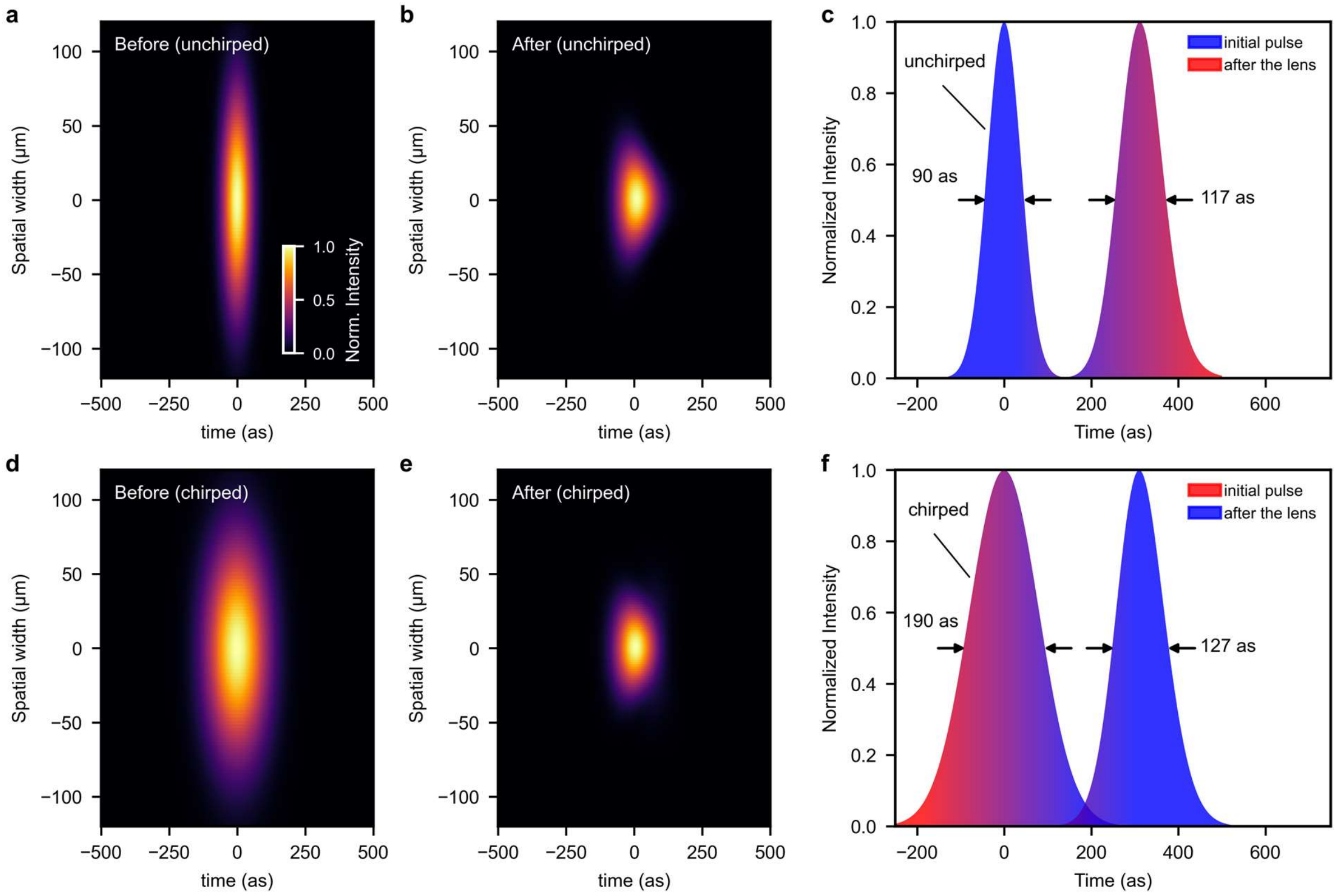

**Fig. S10. Simulated spatiotemporal evolution and temporal profiles of attosecond pulses focused by a plasma lens at 200 mbar. a–c,** Simulation results for a transform-limited 90 as pulse centered at 80 eV. **a**, Spatiotemporal intensity before the plasma lens. **b**, After focusing by the plasma lens, the pulse experiences a small stretching to 117 as. **c**, Corresponding temporal envelopes before (blue) and after (purple) the lens. **d–f**, Simulation for an initially atto-chirped 190 as pulse centered at 80 eV. **d**, Spatiotemporal structure before the lens. **e**, After propagation, the pulse is compressed to 127 as. **f**, Temporal envelopes before (purple) and after (blue) the lens.

## S12 Pulse stretching due to group delay dispersion

To quantify the stretching of an attosecond pulse induced by the plasma lens, we consider the effect of group delay dispersion (GDD), defined as:

$$GDD = L\frac{dv_g^{-1}}{d\omega}, \tag{S10}$$

where $L$ is the length of the medium, and $v_g$ is the group velocity. For a Gaussian transform-limited pulse with an initial FWHM intensity duration $\tau_0$, the resulting pulse duration after experiencing GDD is given by:

$$\tau = \tau_0 \sqrt{1 + \left(\frac{4 \ln 2 \cdot GDD}{\tau_0^2}\right)^2}. \tag{S11}$$

For plasma, $v_g$ can be determined using Eq. (1). In the XUV regime, where the frequency is much larger than the plasma frequency for typical experimental conditions, $\omega \gg \omega_p$, the group velocity can be approximated as $v_g \approx cn$, with $c$ being the speed of light in a vacuum[9].

For the conditions described in Fig. S10**a**, we find that a transform-limited 90 as pulse passing through a plasma lens would experience -2140 as$^2$ of GDD and be elongated to 111 as, making GDD the main contributing factor to pulse stretching in this case.

## S13 Analytical formula for the effective focal length of the plasma lens

Starting with a semi-empirical plasma density model described in Section S6, it can be shown that an analytical expression for the effective focal length $f$ of the plasma lens equates to:

$$f = \frac{\omega^2}{\omega_{p0}^2} \frac{3r_{ch}^2}{L}, \tag{S12}$$

where $\omega_{p0}^2$ is the on-axis plasma frequency that depends on the on-axis free-electron density as described by Eq. (S2). The above expression, together with Eq. (S2), allows one to determine the effective focal length of the lens that depends on the initial gas density inside the capillary.

When diffraction of the beam cannot be neglected, a more accurate focal position after the lens may be found using Gaussian optics. For a lens located at $z = 0$, a Gaussian beam will be focused at distance $z_f$ after the lens:

$$z_f = \frac{f}{1 + \left(\frac{f}{z_R}\right)^2 \left(1 - \frac{z_0}{f}\right)^2}, \tag{S13}$$

where $z_R$ is the Rayleigh length of the beam before the lens, and $z_0$ is the distance between the beam waist and the lens prior to focusing. Similar equations have been derived by Hubbard et al.[15] to describe the focusing of intense laser fields by thin plasma channels when relativistic effects are absent.

## S14 Beam waist measurement of the attosecond pulse centered at 20 eV far from the focus

An additional knife-edge measurement performed for the attosecond pulse centered at 20 eV, with beam radius values taken far from the focal plane, is shown in Fig. S11. Unlike the measurement at 20 cm after the plasma lens, where the presence of plasma reduced the beam radius from 97±2 µm to 40±2 µm (see also Fig. 2**a**), this additional far-field measurement shows the opposite effect: the beam radius increases in the presence of plasma, consistent with the behavior of a focused Gaussian beam beyond its waist. Additionally, the measured values exhibit good agreement with calculations both from the numerical simulations using the WPM and the analytical model described above.

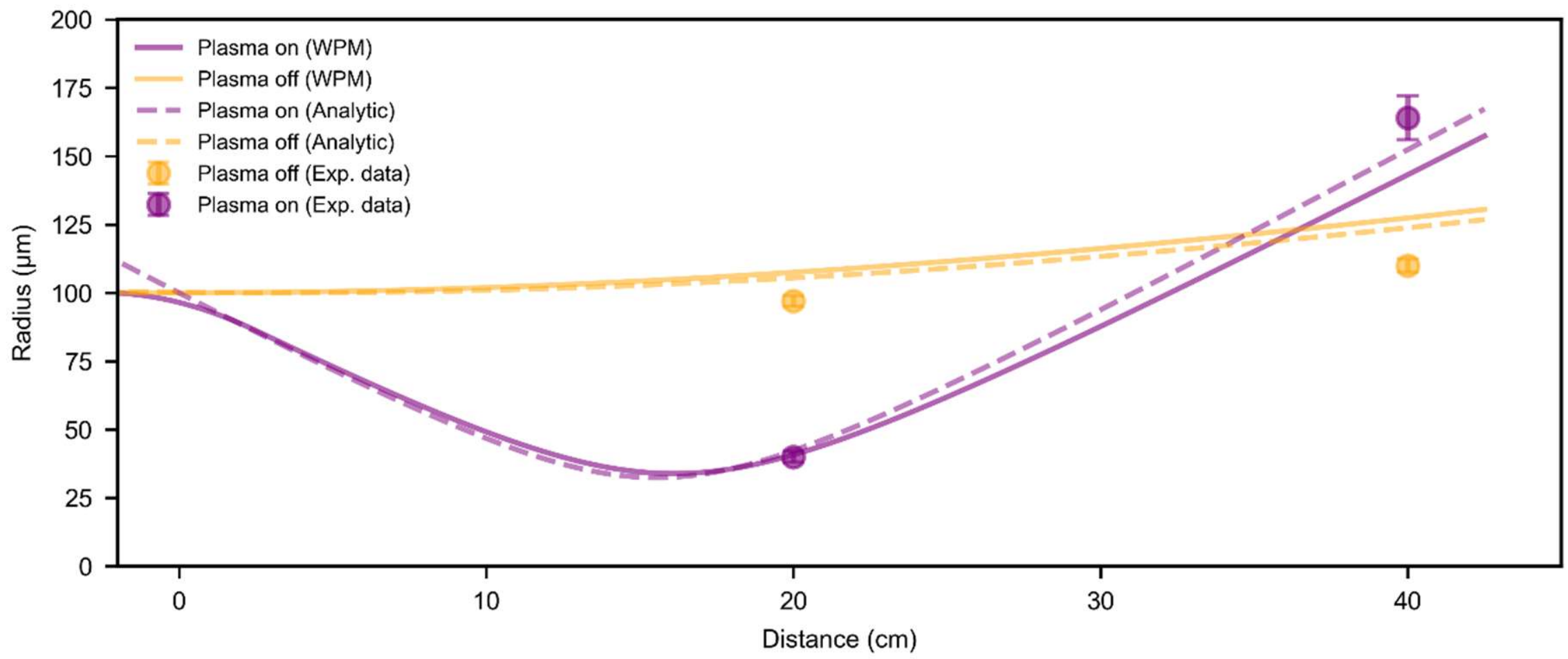

**Fig. S11. Evolution of the beam radius for a 20 eV attosecond pulse after propagation through the plasma lens.** Solid lines show numerical results using the WPM for an empty capillary (yellow) and a fully ionized hydrogen-filled capillary at 63 mbar (purple). Dashed lines represent corresponding analytical predictions. Experimental beam radii measured at 20 cm and 40 cm after the lens are shown as circles; each value is obtained from a single knife-edge scan, with the central point given by the extracted radius from a Gaussian fit and the error bars representing the standard deviation from the fit. Other experimental conditions are identical to those in Fig. 2 of the main text.